\documentclass[graphicx,preprint]{revtex4-2}

\usepackage{amsmath}
\usepackage[british]{babel}
\usepackage{graphicx}
\usepackage{amssymb}
\usepackage{amsthm}
 \usepackage{tikz}
 \usetikzlibrary{tikzmark}

\bibstyle{aipnum4-2}

\graphicspath{{figures/}}

\newcommand{\mx}[1]{\ifmmode\mathbf{#1}\else\textbf{#1}\fi}

\newcommand{\figwidth}{0.95\textwidth}

\newcommand{\labelfont}{\fontsize{10}{10}\selectfont}
\newcommand{\apanel}[1]{\textbf{\textit{\labelfont  #1}}}

\newsavebox{\figbox}
\newcommand{\fig}[1]{figure~\ref{fig:#1}}

\allowdisplaybreaks
\begin{document}
\title{Beyond Michaelis-Menten: Allosteric Rate Control of Chaos}

\author{Tjeerd V. olde Scheper}
\affiliation{School of Engineering, Computing and Mathematics, Oxford Brookes University\\Wheatley Campus, Oxford, OX33 1HX, United Kingdom\\ \url{tvolde-scheper@brookes.ac.uk}}
\begin{abstract}
	The method developed by Michaelis and Menten was foundational in the development of our understanding of biochemical reaction kinetics. 
	Extended models of metabolism encapsulated by reaction rate theory, stochastic reaction models, and dynamic flux estimation, amongst others, address aspects of this fundamental idea. The limitations of these approaches are well understood, and efforts to overcome those issues so far have been plentiful but with limited success. The known issues can be summarised as the sole dependent relation with substrate concentration, the encapsulation of rate in a single relevant scalar, and the subsequent lack of functional control that results from this assumption.
The Rate Control of Chaos (RCC) is a nonlinear control method that has been shown to be effective in controlling the dynamic state of biological oscillators based on the concept of rate limitation of the exponential growth in chaotic systems.
 Extending RCC with allosteric properties allows robust control of the enzymatic process, and replicates the Michaelis-Menten kinetics. The emergent dynamics is robust to perturbations and noise but susceptible to regulatory adjustments.  This control method adapts the control parameters dynamically in the presence of a ligand, and permits introduction of energy relations into the control function.  The dynamic nature of the control eliminates the steady-state requirements and allows the modelling of large-scale dynamic behaviour, potentially addressing issues in metabolic disorder and failure of metabolic control.
\end{abstract}
\maketitle
\pagebreak

\section{Introduction}
\label{sec:intro}
The development of methods to improve the understanding of metabolic control mechanisms is currently of great interest again with significant developments in the area of machine learning applied to biophysics \cite{AlQuraishi.2019}. The potential of such approaches to elucidate the conformational properties of protein structures can be said to greatly accelerate the development of a more complete understanding of the functional aspects of protein interactions. Such a development should feed directly into the control of reaction rates, allowing a functional relation between the biophysical properties and the phenotypical dynamics to be investigated. This will require further progression of the interpretation of the enzymatic control properties that is consistent with the known relations and can be extended towards biophysical properties. The Allosteric Rate Control of Chaos (ARCC) method of nonlinear control allows such an interpretation to be applied and further developed.

The method of reaction rate estimation developed by Michaelis and Menten \cite{Michaelis.1913} has been foundational in the development of our understanding of reaction kinetics. Subsequent experimental evidence has supported the concept to drive further modelling results \cite{Monod.1965}. Extended models of metabolism have been encapsulated by reaction rate theory \cite{Truhlar.1996}, stochastic reaction models \cite{Liao.2015nvl}, and dynamic flux estimation \cite{Pollak.2005} for example, addressing aspects of this fundamental idea. Although the limitations of these approaches are well understood, and efforts to overcome those issues have been plentiful, they have met with somewhat limited success \cite{Nussinov.2015,Truhlar.1996,Hill.1977mw,Garcia-viloca.2004,Gunasekaran.2004}. 
	
For the purpose of this work, the essential issues with the standard model approaches can be summarised as follows \cite{Voit.2017}. Firstly, the sole dependent relation of the reaction rate equation with substrate concentration. Secondly, the encapsulation of the reaction rate in a single relevant scalar. And thirdly, arguably the most important aspect, the subsequent lack of functional control that results from these assumptions. Methods to address the absence of control methodology included in the reaction rate paradigm have been attempted, such as the well known Metabolic Control Analysis  (MCA) \cite{Fell.1997,Hofmeyr.1997} that describes the distribution of control of the flux in a reaction cascade. As the control coefficients are linear approximations, this imposes great restrictions on the response to perturbations and requires a steady-state assumption throughout \cite{Schuster.1999,Hatzimanikatis.1998}. 
		
A suitable methodology for kinetic models, that is not constrained by the issues noted earlier, would require four core elements. Firstly, a demonstrable application within the standard control domain to validate existing results \cite{Link.2014}. Secondly, the ability to maintain dynamic stable control \cite{Scheper.2008}. Thirdly, scalability across several orders of magnitude \cite{Savageau.1988}, and lastly allostery, the cooperative binding of ligands that affects reaction rates \cite{Changeux.2013}. The  conformational change of a protein under external influence underlies the mechanism of allostery \cite{Nussinov.2015}, which by itself affects the rate of reaction. The binding of multiple ligands of the same, or different types also affects the reaction rate, often expressed through the Hill coefficient \cite{Hill.1977mw}. These physical interactions of different ligands with the enzyme creates in effect a multidimensional state space for a protein in which its ensemble of subdomains respond to perturbations affecting the reaction rate \cite{Gunasekaran.2004}. These aspects of mutual dependency within a molecular structure are addressed using various methods \cite{Ravera.2015}, with only partial mapping of the binding constraints to reaction rate \cite{Alberty.2006}.
	
It has been shown already that the fractional Michaelis-Menten expression can be readily represented using an exponential function and expressed into a power law  formalism \cite{Savageau.1988,Schuster.1999}. The Rate Control of Chaos  (RCC) is a nonlinear control method that been shown to effectively control the dynamic state of biological oscillators based on the concept of rate limitation of the exponential growth in chaotic systems \cite{Scheper.2017}.  In effect, it changes the rate of the equation such that exponential growth is adjusted, either reduced or enhanced, to allow the variable to remain near a stable orbit or stable state. Combining both these concepts allows robust control of the enzymatic process, and replicates the Michaelis-Menten kinetics, here defined as the Allosteric Rate Control of Chaos (ARCC) method of reaction rate control. Furthermore, the emergent scale-free dynamics can address the issue of scalability by maintaining the system in a critical dynamic state \cite{Scheper.2022}.  The same approach also allows allosteric control by adapting the control parameters dynamically in the presence of a ligand, and makes it possible to introduce energy relations into the rate control function, for example by including the Arrhenius relation.  The dynamic nature of the control removes the steady-state requirements and opens up the opportunity to represent and model large scale controlled networks of dynamic behaviour, potentially addressing issues in metabolic disorders and failure of metabolic control in long-term systemic illnesses.

The selection of a suitable kinetic model that is based on Monod-Wyman-Changeux (MWC) \cite{Monod.1965} kinetics for redefinition within the proposed ARCC paradigm is not a trivial task. Even though many approaches have been made to encapsulate metabolic kinetics, glycolysis, aerobe and anaerobe reduction, and others; most are by definition much reduced from the experimentally established interactions. This is primarily due to the lack of analytical tools available for nonlinear analysis, and linearisation of the models allows in depth understanding of the dynamics and kinetic relations \cite{Voit.2017}. As a consequence, these models are made to exist in some form of steady-state and any emergent dynamics is much reduced. Alternatively, models that are known to show nonlinear unstable dynamics \cite{Olsen.1983} could be controlled, but these may not be considered to address the underlying kinetics sufficiently.  The model used below is based on the Nielsen model \cite{Nielsen.1998} as an initial implementation of the ARCC control with additional extensions.

The proposed mechanism of ARCC requires a conceptional change of view regarding our approach to understanding metabolic control. ARCC has been developed to ensure that the requirements to maintain stability of the underlying dynamics are maintained. This is essential to warrant that the local dynamics of the biochemical interactions remain stable over time, either as a steady-state or a period cycle. The dynamic stability of the reactants is essential for normal physiological function in organisms, and each component within the interplay between catalysts, reactants, enzymes, diffusion, and biophysical adaptation contributes to the remarkable complexity of biochemistry over time.  To demonstrate the utility of the method, it is first shown how the RCC method controls a typical non-biological nonlinear chaotic system. Subsequently, the ARCC method is applied to the  Nielsen metabolic kinetic model, replacing the Michaelis-Menten dynamics \cite{Nielsen.1998}. This is then extended to include a biophysical relation of temperature dependence in the form of the Arrhenius relation. Furthermore, perturbations are added to the ARCC controlled system to show its robustness. Lastly, the allostery property of the ARCC method is demonstrated in a well-known model that is representative of a complex nonlinear inorganic system to show that allostery allows more advanced modes of control \cite{Gyorgyi.1992b}.

\section{Methods}
\label{sec:methods}

\subsection{Rate Control of Chaos}
RCC is a robust nonlinear control method \cite{Scheper.2017} that dynamically adjust the rate of a system in proportion to the total dynamic domain of that system. To illustrate this, the three dimensional model by Rabinovich-Fabrikant (RF) \cite{Rabinovich.1979} was extended with the RCC control. The RF model describes the behaviour of perturbations in dissipative media and is used to describe wave phenomena such as wind waves in water. This model is used to demonstrate the utility of the RCC method of control in a system that is clearly not explicitly devised for rate control and is not based on biochemical considerations. Equations \eqref{eq:rf1}-\eqref{eq:rf3} describe the proportional limit of the three RF variables $x,y$ and $z$. These are then used in equations \eqref{eq:rf4}-\eqref{eq:rf7} to determine the control functions $\sigma_a$, $\sigma_b$, $\sigma_c$, and $\sigma_d$, that control the nonlinear growth terms in the RF equations \eqref{eq:rf8}-\eqref{eq:rf10}.

\begin{align}
	\label{eq:rf1} q_x&=\frac{x}{(x+\mu_x)}\\
	\label{eq:rf2} q_y&=\frac{y}{(y+\mu_y)}\\
	\label{eq:rf3} q_z&=\frac{z}{(x+\mu_z)}\\
	\label{eq:rf4}  \sigma_a &=f_a \cdot \exp{\left({ \xi_a \cdot q_x \cdot q_y}\right) }\\
	\label{eq:rf5}  \sigma_b &=f_b \cdot \exp\left( \xi_b \cdot q_x \cdot q_z\right) \\
	\label{eq:rf6}  \sigma_c &=f_c \cdot \exp\left( \xi_c \cdot q_x\cdot q_y \cdot q_z\right) \\
	\label{eq:rf7}  \sigma_d &=f_d \cdot \exp\left( \xi_d \cdot q_y \cdot q_z\right) \\
	\label{eq:rf8}  \frac{\partial\,x}{\partial\,t} &=\sigma_d\cdot (y \cdot z)+\sigma_a \cdot (y\cdot x^2)+\gamma \cdot x -y\\
	\label{eq:rf9}  \frac{\partial\,y}{\partial\,t} &=\sigma_b\cdot (3\cdot x\cdot z)+x+\gamma \cdot y-x^3 \\
	\label{eq:rf10}  \frac{\partial\,z}{\partial\,t} &=-(2\cdot\alpha\cdot z)-\sigma_c \cdot (2\cdot x\cdot y\cdot z)
\end{align}
where $\gamma=0.87$, $\alpha=1.1$ and control parameters $\mu_x=\mu_y=\mu_z=4$, $f_a= f_b = f_c = f_d = 1$, $\xi_a=\xi_b=\xi_c=\xi_d=-0.5$. For different values of the control parameters, the system will be controlled into different orbits with respect to the phase space.

\subsection{Nielsen Model}\label{ss:nielsen}

The model proposed by Nielsen, S\o rensen, Hynne and Busse \cite{Nielsen.1998} is a kinetic representation of sustained oscillations in glycolysis in yeast based on experimental measurements. The experimental data was collated from a continuous flow, stirred tank reactor and modelled using MWC kinetics based on earlier work \cite{Termonia.1981wyp}, and extended with flow control. The results of the experimental approach shows that the flow rate is in effect a bifurcation parameter allowing, stable, oscillatory and chaotic dynamics to emerge. A bifurcation parameter may cause a change in dynamic state, making the system stable, or unstable, or chaotic when its value is changed.
Other parameter values were estimated from the experiment or other similar reported results. The equations \eqref{eq:ni1}-\eqref{eq:ni40} describe this model, where the rate equations $v_n, n=[1,25]$ are listed first \eqref{eq:ni1}-\eqref{eq:ni25} and labelled in order. The differential equations \eqref{eq:ni26}-\eqref{eq:ni40} describe the reactants fractions based on these rates. Note that $v_{16}$, $v_{17}$, $v_{19}$, $v_{21}$, and $v_{22}$ contain the relevant Michaelis-Menten kinetics. Other rates are either flow dependent, constant, or simply linear. 

\begin{align}
\label{eq:ni1} 	v_1&=(3.5-ATP) \cdot flow\\
\label{eq:ni2} v_2&=(1.1-ADP) \cdot flow\\
\label{eq:ni3 }v_3&=(0.24-NADH) \cdot flow\\
\label{eq:ni4} v_4&=(4-NAD) \cdot flow\\
\label{eq:ni5} v_5&=(50-GLC) \cdot flow\\
\label{eq:ni6} v_6&=F6P \cdot flow\\
\label{eq:ni7} v_7&=FBP \cdot flow\\
\label{eq:ni8} v_8&=GAP \cdot flow\\
\label{eq:ni9} v_9&=DPG \cdot flow\\
\label{eq:ni10} v_{10}&=PEP\cdot flow\\
\label{eq:ni11} v_{11}&=PYR \cdot flow\\
\label{eq:ni12} v_{12}&=ACA \cdot flow\\
\label{eq:ni13} v_{13}&=EtOH \cdot flow\\
\label{eq:ni14} v_{14}&=AMP \cdot flow\\
\label{eq:ni15} v_{15}&=P  \cdot flow\\
\label{eq:ni16} v_{16}&=\frac{V_1  \cdot ATP \cdot GLC}{(K_{1GLC}+GLC) \cdot(K_{1ATP}+ATP)}\\
\label{eq:ni17} v_{17}&=\frac{V_2  \cdot {F6P}^2  \cdot ATP}{\left(K_2  \cdot \left( 1+ k_2  \cdot \left(\frac{ATP}{AMP}\right)^2\right)+{F6P}^2 \right)  \cdot (K_{2ATP}+ATP)}\\
\label{eq:ni18} v_{18}&=k_{3f}  \cdot FBP-k_{3b}  \cdot {GAP}^2\\
\label{eq:ni19} v_{19}&=\frac{V_4  \cdot GAP  \cdot NAD}{(K_{4GAP}+GAP)  \cdot (K_{4NAD}+NAD)}\\
\label{eq:ni20} v_{20}&=k_{5f}  \cdot DPG  \cdot ADP-k_{5b}  \cdot PEP  \cdot ATP\\
\label{eq:ni21} v_{21}&=\frac{V_6  \cdot PEP  \cdot ADP}{(K_{6PEP}+PEP)  \cdot (K_{6ADP}+ADP)}\\
\label{eq:ni22} v_{22}&=\frac{V_7  \cdot PYR}{(K_{7PYR}+PYR)}\\
\label{eq:ni23} v_{23}&=k_{8f}  \cdot ACA  \cdot NADH- k_{8b}  \cdot EtOH  \cdot NAD\\
\label{eq:ni24} v_{24}&=k_{9f}  \cdot AMP  \cdot ATP- k_{9b}  \cdot {ADP}^2\\
\label{eq:ni25} v_{25}&=k_{10}  \cdot F6P\\
\label{eq:ni26} \frac{\partial\,ATP}{\partial\,t}&=v_1+v_{20}+v_{21}-v_{16}-v_{17}-v_{24}\\
\label{eq:ni27} \frac{\partial\,ADP}{\partial\,t}&=v_2+v_{16}+v_{17}+2  \cdot v_{24}-v_{20}-v_{21}\\
\label{eq:ni28} \frac{\partial\,AMP}{\partial\,t}&=v_{14}-v_{24}\\
\label{eq:ni29} \frac{\partial\,GLC}{\partial\,t}&=v_5-v_{16}\\
\label{eq:ni30} \frac{\partial\,F6P}{\partial\,t}&=v_{16}-v_6-v_{17}-v_{25}\\
\label{eq:ni31} \frac{\partial\,FBP}{\partial\,t}&=v_{17}-v_7-v_{18}\\
\label{eq:ni32} \frac{\partial\,GAP}{\partial\,t}&=2  \cdot v_{18}-v_8-v_{19}\\
\label{eq:ni33} \frac{\partial\,NAD}{\partial\,t}&=v_4+v_{23}-v_{19}\\
\label{eq:ni34} \frac{\partial\,NADH}{\partial\,t}&=v_3+v_{19}-v_{23}\\
\label{eq:ni35} \frac{\partial\,DPG}{\partial\,t}&=v_{19}-v_9-v_{20}\\
\label{eq:ni36} \frac{\partial\,PEP}{\partial\,t}&=v_{20}-v_{10}-v_{21}\\
\label{eq:ni37} \frac{\partial\,PYR}{\partial\,t}&=v_{21}-v_{11}-v_{22}\\
\label{eq:ni38} \frac{\partial\,ACA}{\partial\,t}&=v_{22}-v_{12}-v_{23}\\
\label{eq:ni39} \frac{\partial\,EtOH}{\partial\,t}&=v_{23}-v_{13}\\
\label{eq:ni40} \frac{\partial\,P}{\partial\,t}&=v_{25}-v_{15}
\end{align}
where $V_1=0.5,  K_{1GLC}=0.1,  K_{1ATP}=0.063, V_2=1.5, K_2=0.002,  k_ 2=0.017,  K_{2ATP}=0.01,  k_{3f}=1, k_{3b}=50,  V_4=20,  K_{4GAP}=1, K_{4NAD}=1,  k_{5f}=1,  k_{5b}=0.5,  V_6=10, K_{6PEP}=0.2,  K_{6ADP}=0.3, V_7=2,  K_{7PYR}=0.3,  k_{8f}=1,  k_{8b}=0.000143, k_{9f}=10, k_{9b}=10, k_{10}=0.05$. The parameter that represents the flow rate is maintained at a constant value, $flow=0.0082$, but this is a bifurcation parameter changing the dynamic response when adjusted.

The initial values of the ion concentrations are $ATP=4.49064, ADP=0.108367, AMP=0.00261149,  GLC=0.0112817, F6P=0.65939,  FBP=0.00770135, GAP=0.00190919, NAD=3.62057,  NADH=0.616118, DPG=0.299109, PEP=0.0021125, PYR=0.00422702, ACA=0.0738334, EtOH=0.33981, P=0$.

\subsection{ARCC}
Given the representative model described in section \ref{ss:nielsen}, the application of the Allosteric Rate Control of Chaos, can be shown as follows. Firstly, the rate equations that are not of the Michaelis-Menten type can be left as they are, although it could be argued that these are also rate controlled. This will not be explored further for the moment. Secondly, the five rate equations that are based on the proportional representation are defined as the rate control applied to the reactants as normal, but scaled using the RCC method.

To create suitable representations of the ARCC control function for this specific application of the control method, RCC is generalised as

\begin{align}
	\label{eq:grcc1} q(X_k)&=\frac{X_k^n}{X_k^n+\mu_X^n}\\
	\label{eq:grcc2} \sigma_k(X_k)&=f_k \cdot\exp{\left(\xi\cdot q(X_k)+\theta_k\right)} 
\end{align}
where $q(X_k)$ is the quotient of the reactant $X$ for sample $k$, $n$ is the Hill coefficient applied to the RCC control function, and is normally taken as $n=1$. Other values for this coefficient will be explored in the subsequent section \ref{ss:ac}. The control coefficient $\mu_X^k$ is the maximum range of the reactant greater or equal to $\max(X)$ for sample $k$. $f_k$ is the control scalar, normally $f_k=1$, and the bias term $\theta_k=0$ for most cases. In section \ref{ss:arr}, its functional role as biophysical regulator term will be explored.

Equation \eqref{eq:arrc1} describes the reactants under consideration within these rate equations conform \eqref{eq:grcc1}, where $X\in \{GLC,ATP,ADP,AMP,F6P,GAP,NAD,PEP\}$ and $\mu_X=5$ for all reactants. The control functions $\sigma_X$ described in \eqref{eq:arrc2}-\eqref{eq:arrc6} is determined by the same considerations as the original model \cite{Nielsen.1998}, where the argument of the exponent is defined by the reactants' rates as they contribute to the reaction.

\begin{align}
\label{eq:arrc1} q_X &=\frac{X}{(X+\mu_X)}\\
\label{eq:arrc2} \sigma_{v_{16}}&=f_{16}\cdot \exp{\left( \xi_{16}\cdot  q_{GLC}\cdot q_{ATP}+\theta_{16}\right) }\\
\label{eq:arrc3} \sigma_{v_{17}}&=f_{17}\cdot \exp{\left( \xi_{17}\cdot q_{F6P} \cdot q_{ATP} \cdot \left( \frac{ q_{ATP} }{q_{AMP}} \right)^{n_{17}}+\theta_{17}\right)}\\
\label{eq:arrc4} \sigma_{v_{19}}&=f_{19}\cdot \exp{\left( \xi_{19}\cdot q_{GAP} \cdot q_{NAD}+\theta_{19} \right)}\\
\label{eq:arrc5} \sigma_{v_{21}}&=f_{21}\cdot \exp{\left( \xi_{21} \cdot q_{PEP}\cdot q_{ADP}+ \theta_{21}\right) }\\
\label{eq:arrc6} \sigma_{v_{22}}&=f_{22}\cdot \exp{\left( \xi_{22} \cdot q_{P}+\theta_{22}\right)}\\
\label{eq:arrc7} v_{16}&=V_1\cdot \sigma_{v_{16}} \cdot ATP \cdot GLC\\
\label{eq:arrc8} v_{17}&=V_2\cdot \sigma_{v_{17}} \cdot {F6P}^2\cdot ATP\\
\label{eq:arrc9} v_{19}&=V_4\cdot \sigma_{v_{19}}\cdot GAP\cdot NAD\\
\label{eq:arrc10} v_{21}&=V_6\cdot \sigma_{v_{21}}\cdot PEP \cdot ADP\\
\label{eq:arrc11} v_{22}&=V_7\cdot \sigma_{v_{22}}\cdot PYR
\end{align}
where $f_{16}=1, \xi_{16}=-3,  \theta_{16}=0, f_{17}=1, \xi_{17}=-1.5, \theta_{17}=0, f_{19}=1,  \xi_{19}=-3, \theta_{19}=0, f_{21}=1, \xi_{21}=-3, \theta_{21}=0, f_{22}=1, \xi_{22}=-3, \theta_{22}=0$ and $n_{17}=2$. All other equations, and parameter values, are the same as in the model described above in \eqref{eq:ni1}-\eqref{eq:ni40}. The implementation of the Nielsen model was derived from the paper and from the repository on BioModels \cite{BioModels2020} for this model. The representation of ARCC is based on the provided information derived from these publications, and not yet matched against experimental data. All subsequent results are therefore not labelled as specific concentrations, but as resulted amounts of each reactant against arbitrary time.

\subsection{Temperature Dependence}\label{ss:arr}

The use of the Allosteric Rate Control of Chaos as a means of controlling the reaction rate also allows the inclusion of the temperature dependent reaction rates. The reaction rate as a function of the operating temperature is often of specific concern in conditions not generally applicable to biological physiology, but it makes an important contribution to stable dynamics.  Different conditions and mechanisms have been proposed for encapsulating reaction temperature in reactions \cite{Carvalho-Silva.2019}. The general case can be describe by the original contribution of Arrhenius based on a exponential decay function:

\begin{align}
	k=A\cdot\exp{\left({\frac{-E_a}{R\cdot T}}\right)}
\end{align}
where $R$ is the gas constant, and $T$ the absolute temperature. The pre-factor $A$ is often temperature independent, and is sometimes given the name of a frequency factor. $E_a$ is the energy of activation representing the threshold of energy needed for a molecule to contribute to the reaction. This function can be included into the ARCC model as the contribution condition represented by the bias term $\theta$, in effect modifying the control conditions in proportion to the temperature of the reaction. Inclusion of the Arrhenius function within the RCC control function will cause the control to be greater at lower temperature, which reflects the effort needed to stabilise the reaction dynamics. This property is under further investigation.

The equations \eqref{eq:arrc2}-\eqref{eq:arrc6} have been adjusted to include the Arrhenius relation as follows:

\begin{align}
\theta_{16}&=- \frac{E_{16}}{R\cdot T_{abs}}\\
\theta_{17}&=- \frac{E_{17}}{R\cdot T_{abs}}\\
\theta_{19}&=- \frac{E_{19}}{R\cdot T_{abs}}\\
\theta_{21}&=-\frac{E_{21}}{R\cdot T_{abs}}\\
\theta_{22}&=-\frac{E_{22}}{R\cdot T_{abs}}
\end{align}
where $ E_{16}=3.275, E_{17}=3275, E_{19}=3.275, E_{21}=3.275, E_{22}=3.275,
 R=8.31446261815324$, the gas constant, and temperature $T_{abs}=310.15$ (approx. 37\textdegree C). The values for the activation energy $E$ where estimated by matching the period of the original model, note that the control scalar $f_{17}=0.175$ for these models. Other values for the activation energy given a specific temperature can readily be found to match a specific oscillation pattern. It should be clear that inclusion of the Arrhenius equation in the ARCC model, does not preclude other more modern biophysical approaches to be applied to the reaction rate \cite{Carvalho-Silva.2019}.

\subsection{Allosteric Control}\label{ss:ac}
To include allostery as a means of control within the Rate Control of Chaos method, the generalised form of RCC is applied to the Belousov-Zhabotinsky reaction. This demonstrates that the addition of allostery can greatly affect the stability and dynamic response as has been shown to be the case in biochemical reactions. The system has already been shown to be controllable using RCC \cite{Scheper.2017} and can employ local control to stabilise a spatiotemporal array of oscillators. This simulation is replicated here, and then extended by changing the diffusion rate to cause spatiotemporal instabilities in the controlled chaotic system. The  $n$-parameter in equation \eqref{eq:bz1} is subsequently adjusted to simulate the effect of cooperative binding and thereby affect the dynamic response of the network of oscillators. The RCC controlled BZ network is modelled as a one dimensional array of oscillators with diffusion between each subsequent oscillator, and with the first and last  oscillator to diffuse their products to each other. This will avoid potential edge effects in the one-dimensional array that may cause further instabilities. The resulting equations are as follows:

\begin{align}
\label{eq:bz1} \sigma_k(x_k)&=\exp{\left(\frac{\xi\cdot x_k^n}{x_k^n+\mu^n}\right)}\\
\label{eq:bz2} y_k&=\left(\frac{\alpha\cdot  k_6\cdot  Z_0\cdot  V_0\cdot  z_1\cdot  v_1}{k_1\cdot  H\cdot  X_0\cdot  x_1+k_2\cdot  A\cdot  H^2 + k_{fk}}\right)/Y_0\\
\label{eq:bz3} \begin{split}\frac{\partial x_k}{\partial t}&=T_0\cdot  (-k_1\cdot  H\cdot  Y_0\cdot  x_1\cdot  y_1 +k_2\cdot  A\cdot  H^2\cdot  (Y_0/X_0)\cdot  y_k\\&-2\cdot  k_3\cdot  X_0\cdot  x_1^2 +0.5\cdot  k_4\cdot  A^{0.5}\cdot  H^{1.5}\cdot  X_0^{-0.5}\cdot  (C-Z_0\cdot  z_k)\cdot  x_k^{0.5}\cdot  \sigma_k(x_k)\\&-0.5\cdot  k_5\cdot  Z_0\cdot  x_k\cdot  z_k -k_f\cdot  x_k )\end{split}\\
\label{eq:bz4} \begin{split}\frac{\partial z_k}{\partial t}&=T_0\cdot(k_4\cdot A^{0.5}\cdot H^{1.5}\cdot X_0^{0.5}\cdot(C/Z_0-z_k)\cdot x_k^{0.5}\cdot\sigma_k(x_k)\\& -k_5\cdot X_0\cdot x_k\cdot z_k - a\cdot k_6\cdot V_0\cdot z_k\cdot v_k - b\cdot k_7\cdot M\cdot z_k- k_f\cdot z_k)\end{split}\\
\label{eq:bz5} \begin{split}\frac{\partial v_k}{\partial t}&=T_0\cdot(2\cdot k_1\cdot H \cdot X_0 \cdot (Y_0/V_0)\cdot x_k\cdot y_k +k_2\cdot A\cdot H^2\cdot (Y_0/V_0)\cdot y_k\\&+k_3\cdot (X_0^2/V_0)\cdot x_k^2\cdot \sigma_k(x_k)-a\cdot k_6\cdot Z_0\cdot z_k\cdot v_k-k_f\cdot v_k)\end{split}
\end{align}
where $\sigma_k(x_k)$ is the RCC function and the subscript $k\in\{1,\ldots,10\}$ indicates the index of each BZ reaction in 1D space (for the BZ chaotic system parameter values, see \cite{Gyorgyi.1992, Gyorgyi.1992b}). The bifurcation parameters $k_f$ were chosen differently for each oscillator. Values were $k_{f1}=0.003, k_{f2}=0.00032,k_{f3}=0.00034,k_{f4}=0.00036,k_{f5}=0.00038,k_{f6}=0.0004,k_{f7}=0.00031,k_{f8}=0.00035,k_{f9}=0.00039,k_{f10}=0.00039 $ throughout the experiments for the used network models. The diffusion was modelled using finite difference for all variables where each diffusion rate constant $D_k \in [0,1]$ was set to the same value for each individual BZ system.

\subsection{Modelling}
The equations were modelled using the EuNeurone numerical integration software, which is available on Zenodo \citep{Scheper.2013}. The results can nevertheless be readily simulated using other fixed-step numerical integration tools. The numerical integrators used were the standard fixed step integration Runge-Kutta RK4,  and Fehlberg-RK algorithms \citep[pages 363-366]{Gough.2009}. Results were then exported into Hierarchical Data Format  5 (HDF5), subsequently analysed, and plotted using Matlab. For each model in the manuscript, a corresponding file is available from Zenodo \citep{Scheper.2023df} that allows reconstruction of the results.

\section{Results}

\subsection{Rate Control of Chaos Results}
\begin{figure*}[ht]
\begin{minipage}{\figwidth}
\includegraphics[width=\textwidth]{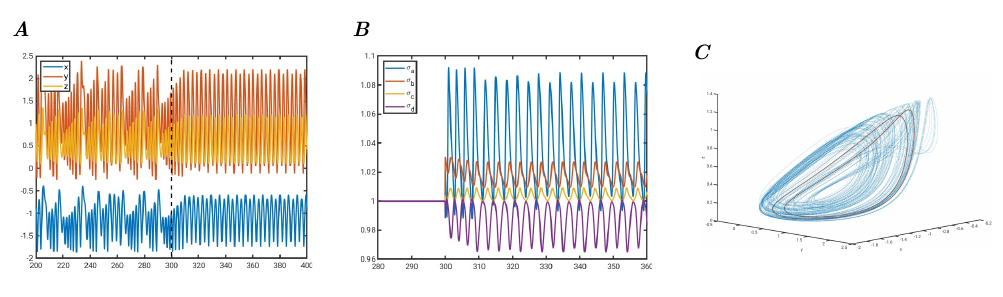}
\end{minipage}
\caption{\label{fig:1} Rate Control of Chaos applied to the Rabinovich-Fabrikant equations. \apanel{A} Evolution in time of the three variable with the RCC control enabled at the dotted line, showing chaos before and stable two orbit afterwards. \apanel{B} The four control functions in time showing the variability to rate limit the variables in time. \apanel{C} Phase space plot in 3D of the three variables, showing in blue the chaotic state, and in red superimposed the controlled 2 orbit. }
\end{figure*}

As a demonstration of the effectiveness of the RCC method to control nonlinear dynamic systems, the method is applied to a nonlinear chaotic Rabinovich-Fabrikant  system shown in \fig{1}. Further applications of the principle may be found in the referenced sources \cite{Scheper.2017,Scheper.2017z9f}. The behaviour shown is that the unstable chaotic dynamics is controlled into a periodic oscillation with a two orbit. The first panel shows that when the control is enabled, the system will converge to the orbit after a short transition for the three variables in the RF system. The second panel shows the four control functions for each of the variables, that are constant at the value one when disabled but will change in proportion to the oscillation that drives the nonlinearity for every variable. Once control is established and stabilised, the control function will oscillate as well. The last panel shows the difference in a three dimensional phase space, where the three variables are plotted against each other, between chaos and RCC controlled. The blue plot is the chaotic system without control and in red is superimposed the controlled orbit with the transient removed to clearly show the two orbit.

This control of the nonlinear RF model shows how the control will converge to a possible stabilised orbit that is contained within the entire representation space of the three variables. Each variable is locally controlled by the control functions that depend on the proportional growth or shrinkage of each. The immediate result can then be determined as each differential equation is affected by the control to produce the stabilised dynamics over time.

\subsection{Nielsen Model Results}
\begin{figure*}[ht]
\begin{minipage}{\figwidth}
\includegraphics[width=\textwidth]{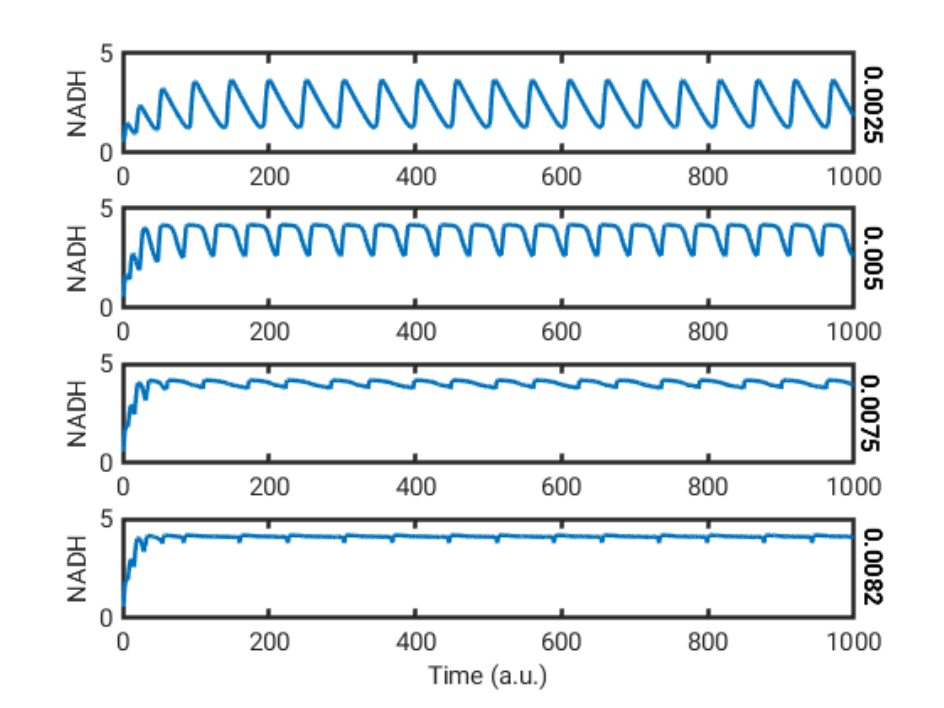}
\end{minipage}
\caption{\label{fig:2} Nielsen model behaviour with Michaelis-Menten kinetics. [NADH]  for different flow amounts (From the top: $0.0025, 0.005, 0.0075, 0.0082$), showing changes in amplitude and wave form.}
\end{figure*}

The original model results are shown in \fig{2} for different values of the bifurcation parameter $flow$. The increase in flow affects the amplitude and wave form of the concentration of the reactants, shown here as the [NADH]. The bottom panel in the figure shows the representative response of the steady-state model for the flow at $0.0082$ for comparison with the original paper.

The change in flow causes a change in amplitude and wave form such that for higher values the  wave becomes more shallow. The parameters in the model were tuned in the original paper to match the experimental results explicitly, and these were used subsequently for the ARCC models as well.

\subsection{ARCC Results}
\begin{figure*}[ht]
\begin{minipage}{\figwidth}
\includegraphics[width=0.9\textwidth]{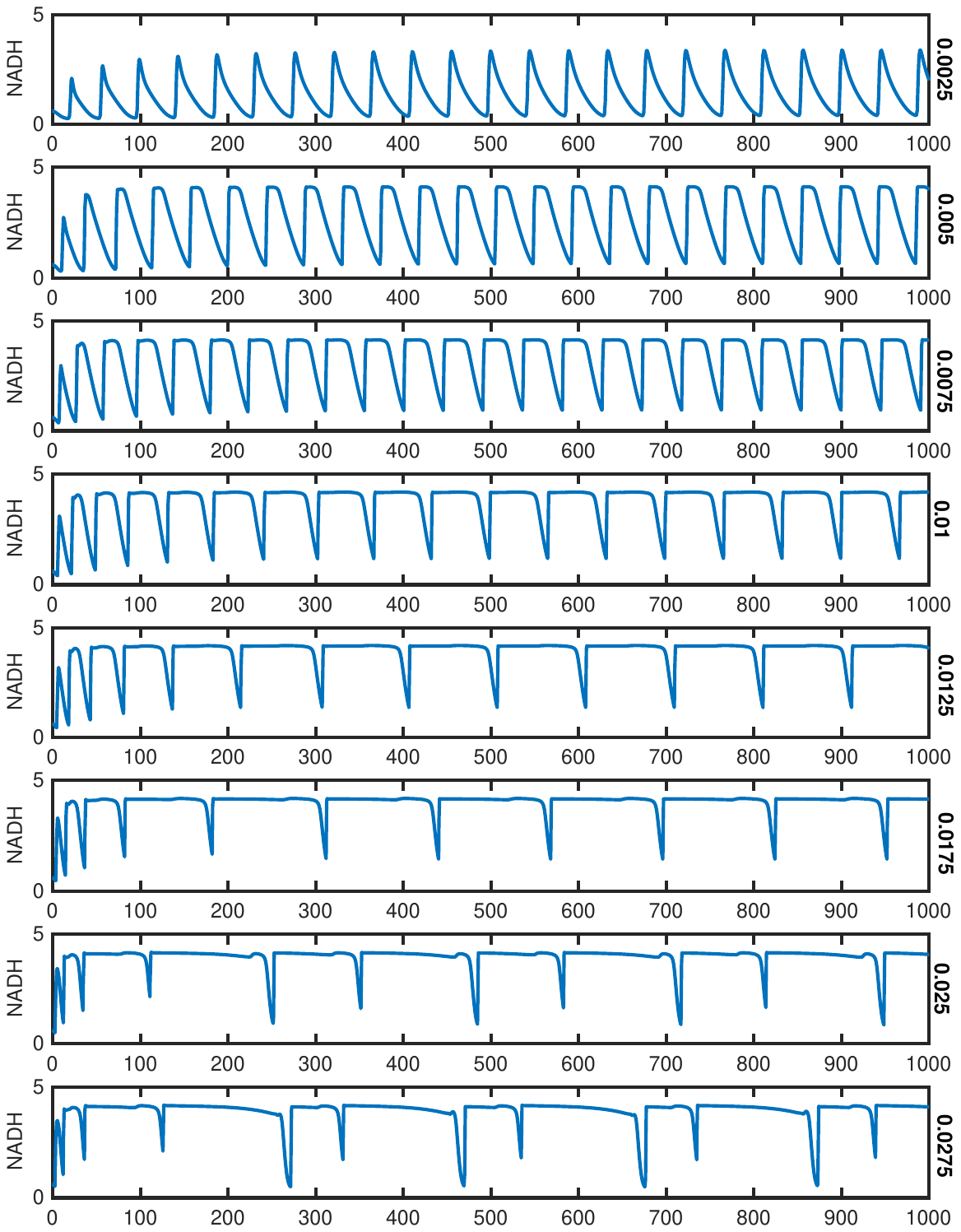}
\end{minipage}
\caption{\label{fig:3} ARCC implementation of the Nielsen model. [NADH]  for different flow amounts (From the top: $0.0025, 0.005, 0.0075, 0.01, 0.0125,0.0175,0.025,0.0275$), showing change in dynamic state and frequency.}
\end{figure*}

Using the approach outlined above, the results of the replacement of the Michaelis-Menten kinetics with ARCC is shown in \fig{3}, where NADH is plotted over time for different values of $flow$. Firstly, it should be noted that the qualitative behaviour is comparable, but somewhat different. The parameters for the kinetic model used in the ARCC version have not been adapted from the original, apart from those that are part of ARCC themselves. This has resulted in a larger range of all variables, for example for NADH as shown, the minimum can be around $0.6$ and the maximum around $4.8$. Secondly, because the control is able to adapt to the changes in flow with different orbits of the reactants, the system is able to cope with much larger increases in flow. The increase in flow also causes changes in shape and form, as well as frequency and in the number of periods, e.g. for a flow above $0.02$, the system shows a two orbit of lower frequency. 

\begin{figure*}[ht]
\begin{minipage}{\figwidth}
\includegraphics[width=\textwidth]{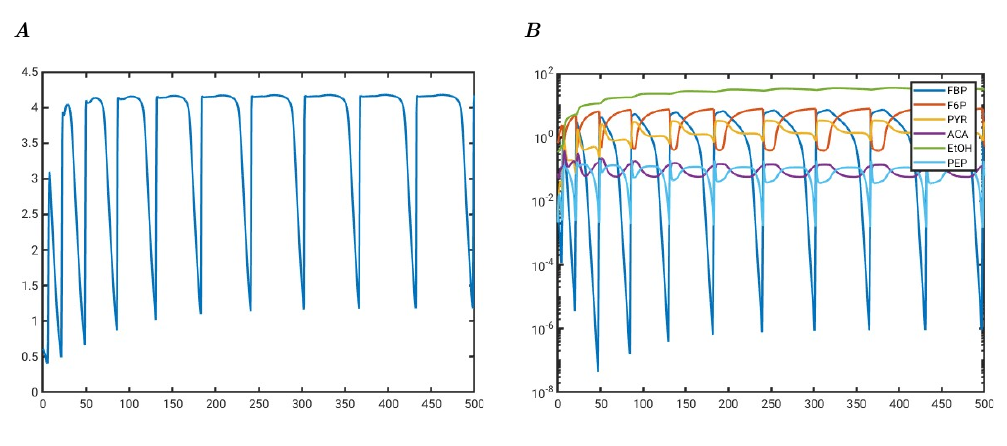}
\end{minipage}
\caption{\label{fig:4} Detail of the ARCC applied to the Nielsen model. \apanel{A} NADH at $flow=0.0082$. \apanel{B} Log-linear plot of amount of reactants over time: FBP (blue), F6P (red), PYR (yellow), ACA (purple), EtOH (Ethanol, green), PEP (light blue).  }
\end{figure*}

In \fig{4} is shown NADH in the left panel at the same flow rate as in the original paper of $0.0082$. The resulting period is similar in shape and frequency but the range is much wider. In the right panel are shown several other reactants in a log-linear plot in time to demonstrate that the entire system is stable oscillating for the given flow. 

\begin{figure*}[ht]
\begin{minipage}{\figwidth}
\includegraphics[width=\textwidth]{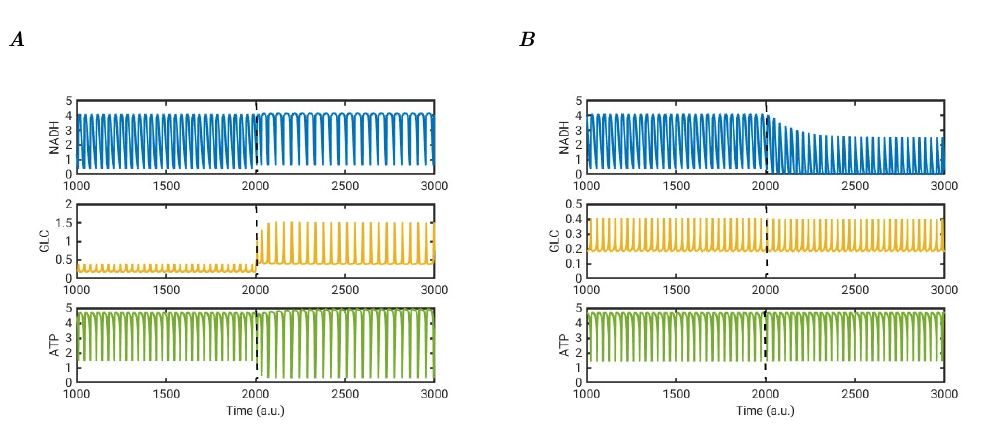}
\end{minipage}
\caption{\label{fig:5} Effect of perturbation on the ARCC model. Shown are the NADH (blue), GLC (yellow), and ATP (green). \apanel{A} Perturbation by an increased supply of GLC ($4.49$) at the midpoint in the plots (indicated by the dotted line). \apanel{B} Perturbation by an increased supply of ACA ($0.3$) at the midpoint (indicated by the dotted line).}
\end{figure*}

 To further explore the behaviour of the ARCC model in comparison to the standard approach, the same model was perturbed by a constant addition of one of two reactants, in a comparable manner as performed by Hynne et al.  \cite{Hynne.2001}, who used the same Nielsen base model. They used the perturbations in an attempt to quench the oscillations to return to steady-state conditions, with some success for the specific model. In the case of the ARCC model, the perturbation results are more interesting as the dynamic response of the system appears to have functional properties. In \fig{5} are shown two perturbations, on the left an increase of $4.49$ of glucose (GLC), and on the right an increase of $0.3$ acetaldehyde (ACA) as a constant added amount.  Both columns show the amount of NADH, GLC and ATP before and after the perturbations at the dotted line. The addition of GLC causes a change in frequency and wave form in the NADH, but importantly, the ATP production is kept consistent as one would expect, although the range of the ATP oscillations has changed. The right panel shows the affect of ACA addition, which reduces the maximal oscillatory values of NADH, but neither GLC consumption nor ATP production seem to be affected, which indicates that the control adjusts correctly for this perturbation and maintains its physiological role independently from this external perturbation.
 
\subsection{Temperature Dependence Results}
\begin{figure*}[ht]
\begin{minipage}{\figwidth}
\includegraphics[width=\textwidth]{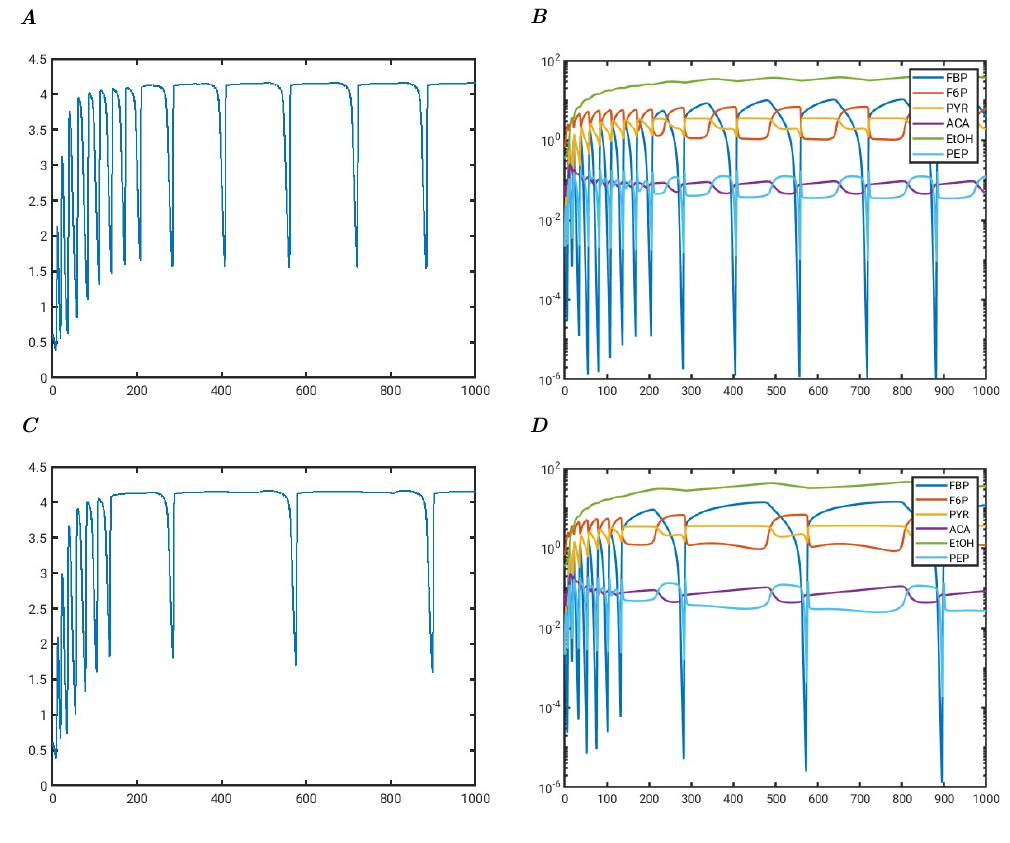}
\end{minipage}
\caption{\label{fig:6} Effect of addition of the Arrhenius relations in the ARCC models. Left column shows the oscillations of NADH for $E_{17}=1500$ \apanel{A}, $E_{17}=2000$ \apanel{B}. The right column shows log-linear plots of amount of reactants over time for the same activation energy as on the left; from top down: FBP (blue), F6P (red), PYR (yellow), ACA (purple), EtOH (green), PEP (light blue).  }
\end{figure*}

The addition of the Arrhenius relation within the ARCC model acts as a constant control adjustment within the rate control function. This addition still acts as a frequency depending function, as the Arrhenius function applies to reaction rates, but is expected to be constant in biophysiological conditions. Additionally, if the activation energy changes in relation to some regulatory mechanism, this also can readily affect the frequency and wave shapes, but will not cause the models to become unstable. 
To illustrate these properties, \fig{6} shows two characteristic oscillations of the model for $E_{17}=1500$, on the top row and $E_{17}=2000$ on the bottom row. The left column in this figure shows the NADH amount and the right column a log-linear plot of several reactants in time. Note that for the higher amount of activation energy, the control function almost doubles the period of the oscillations. The value for the activation energy does not appear to be crucial for these simulations, and further exploration of the functional relation within ARCC is required to elucidate this. It should be clear that biophysical properties affecting the activation energy are expected to be much more complex expressions within the control function, but the ARCC method would allow any such functions to be explored as long as the control function remains within the domain of control. Additionally, if the temperature was changed this would also affect the frequencies of oscillations and for higher temperatures the frequency would increase.

\subsection{Allosteric Control Results}
\begin{figure*}[ht]
\begin{minipage}{\figwidth}
\includegraphics[width=\textwidth]{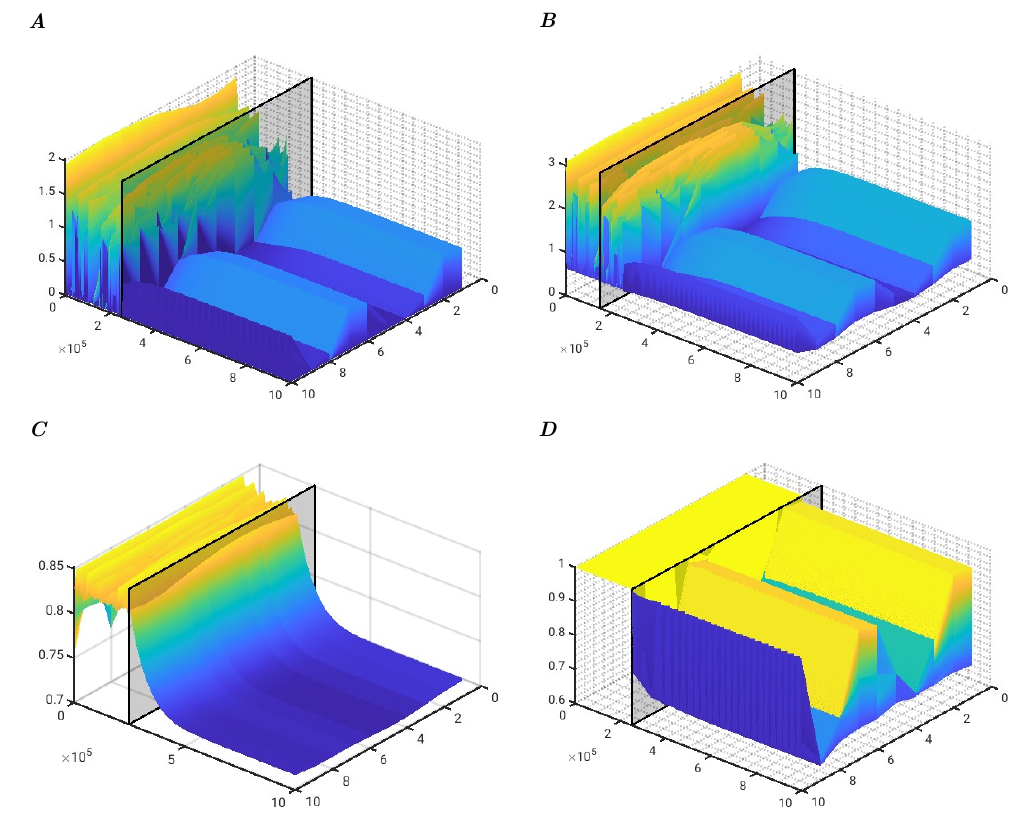}
\end{minipage}
\caption{\label{fig:7} Array of ten  Belousov-Zhabotinsky models with $0.04$ diffusion showing the application of RCC (without allostery) to stabilise the oscillators and prevent spatio-temporal chaos from emerging. The control is enabled at time step $2\cdot 10^5$ indicated by the grey panels. Shown are three of the system variables for each model plotted against time and ordinal position alongside each other. The final panel shows the control function $\sigma$ for each model in time \apanel{A} $x_k, k \in \{1,10\}$, \apanel{B} $z_k, k \in \{1,10\}$, \apanel{C} $v_k, k \in \{1,10\}$, \apanel{D} $\sigma_k, k \in \{1,10\}$ .}
\end{figure*}

The inclusion of allosteric properties within the ARCC model, requires a more extensive model to demonstrate effectively the consequences of allostery within the model. An array of BZ models has already been effectively controlled using the standard RCC method, without allostery, as illustrated in \fig{7} where the oscillations of the ten BZ models is shown for the variables $x, z, v$. Also shown are the control functions for each model in the bottom right panel. The diffusion scalar was chosen to be $0.04$ for this model for which the control shows good results. 

\begin{figure*}[ht]
\begin{minipage}{\figwidth}
\includegraphics[width=\textwidth]{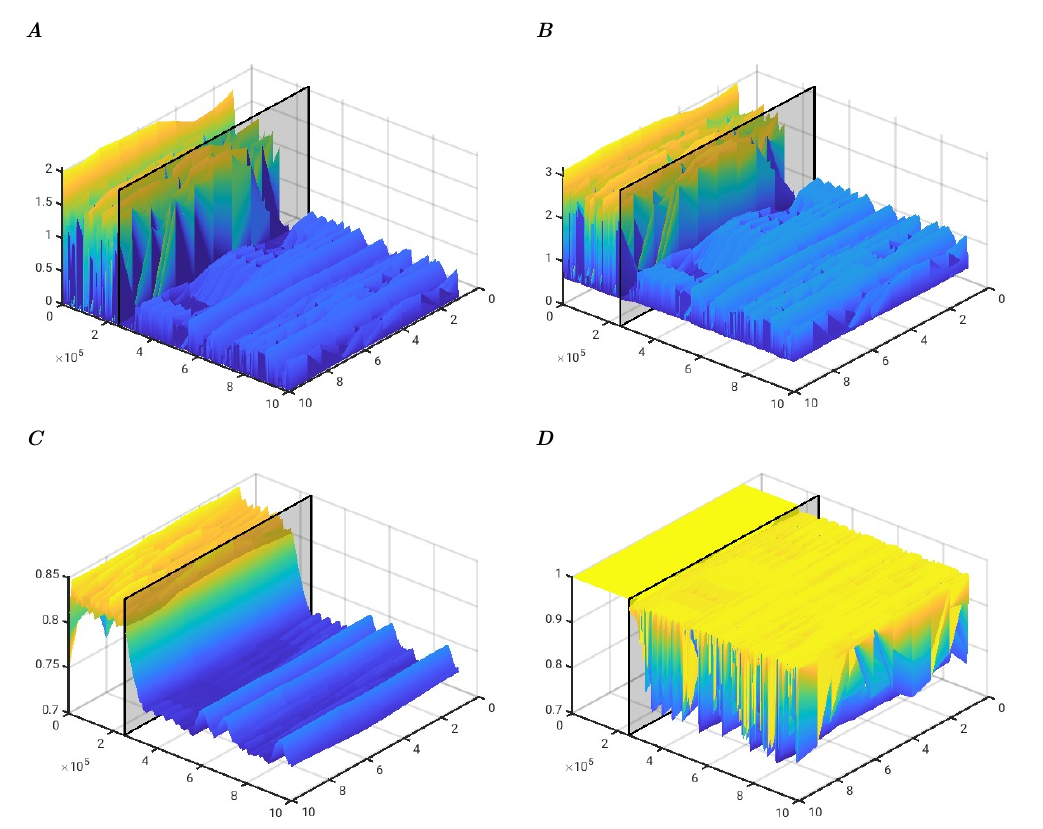}
\end{minipage}
\caption{\label{fig:8} Array of ten  Belousov-Zhabotinsky models with $0.02$ diffusion showing the application of RCC (without allostery) to stabilise the oscillators and failing to prevent spatio-temporal chaos from emerging. The control is enabled at time step $2\cdot 10^5$ indicated by the grey panels. Shown are three of the system variables for each model plotted against time and ordinal position alongside each other. The final panel shows the control function $\sigma$ for each model in time \apanel{A} $x_k, k \in \{1,10\}$, \apanel{B} $z_k, k \in \{1,10\}$, \apanel{C} $v_k, k \in \{1,10\}$, \apanel{D} $\sigma_k, k \in \{1,10\}$ .}
\end{figure*}

However, if the diffusion is reduced below this value, the system can again become unstable, as each oscillator is struggling to remain within a domain of stabilised controlled orbit. This is illustrated in \fig{8}, where the same four system variables of the BZ model in the array of oscillators is shown, as well as the rate control function in the bottom right panel. The variables are still under control and have much reduced amplitudes even though they are clearly not in a steady-state or a periodic orbit. The control functions fluctuate wildly seeking to stabilise each individual oscillator. The parameters in the generic ARCC function \eqref{eq:grcc2} may be adjusted to find a domain of control for each control function.  But it is shown here that adjusting the Hill coefficient $n$ in \eqref{eq:grcc1} is a robust approach to achieve control and permit dynamic adjustment. This should replicate the dynamic changes of metabolic processes by molecular adjustments and signalling molecules.

There are many possible values that may be assigned to the Hill coefficient for this specific set of models, which allows different dynamic behaviours to emerge from the network. To simplify the interpretation of possible properties, two representative examples are shown below. One example for a coefficient value below the value one, and one example with a coefficient value above the value one. In this example, all values have a different value within the range of positive one and positive four.

\begin{figure*}[ht]
\begin{minipage}{\figwidth}
\includegraphics[width=\textwidth]{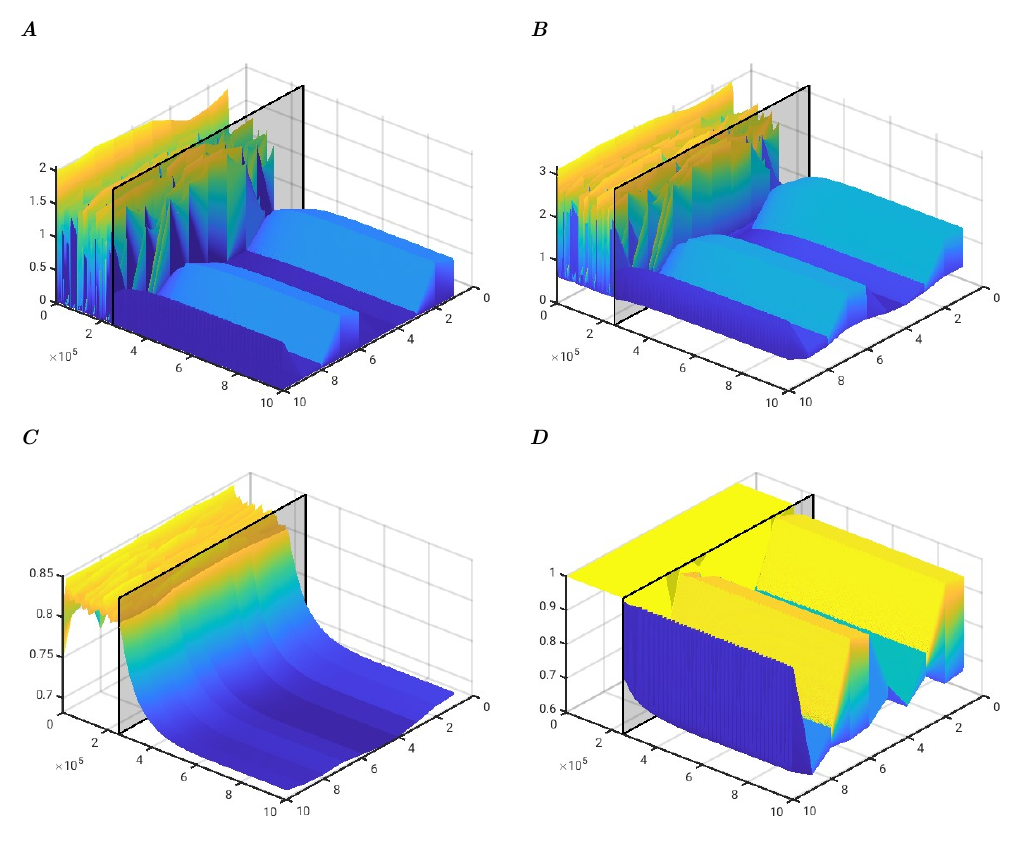}
\end{minipage}
\caption{\label{fig:9} Array of ten  Belousov-Zhabotinsky models with $0.02$ diffusion showing the application of ARCC with allostery values below positive one $[0,1]$ to stabilise the oscillators and succeeding in preventing spatio-temporal chaos from emerging. The control is enabled at time step $2\cdot 10^5$ indicated by the grey panels. Shown are three of the system variables for each model plotted against time and ordinal position alongside each other. The final panel shows the control function $\sigma$ for each model in time \apanel{A} $x_k, k \in \{1,10\}$, \apanel{B} $z_k, k \in \{1,10\}$, \apanel{C} $v_k, k \in \{1,10\}$, \apanel{D} $\sigma_k, k \in \{1,10\}$ .}
\end{figure*}

In the first example of the application of allosteric control to the BZ models, the system returns to stability when $n\leq 1$, as a simulation of competitive binding acts as inhibition of the control on the chaotic dynamics. The example in \fig{9} shows similar behaviour as the original controlled model in \fig{8}. 

\begin{figure*}[ht]
\begin{minipage}{\figwidth}
\includegraphics[width=\textwidth]{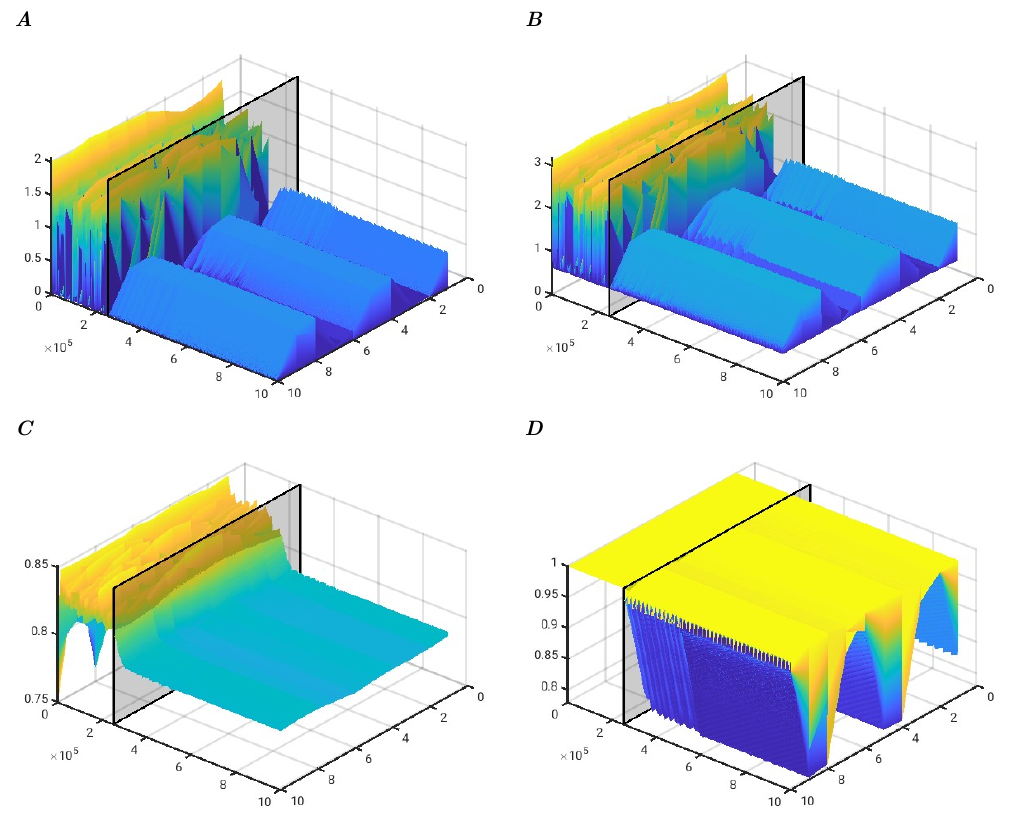}
\end{minipage}
\caption{\label{fig:10} Array of ten  Belousov-Zhabotinsky models with $0.02$ diffusion showing the application of ARCC with allostery values above positive one $[1,4]$ to stabilise the oscillators and succeeding to prevent spatio-temporal chaos from emerging. The control is enabled at time step $2\cdot 10^5$ indicated by the grey panels. Shown are three of the system variables for each model plotted against time and ordinal position alongside each other. The final panel shows the control function $\sigma$ for each model in time \apanel{A} $x_k, k \in \{1,10\}$, \apanel{B} $z_k, k \in \{1,10\}$, \apanel{C} $v_k, k \in \{1,10\}$, \apanel{D} $\sigma_k, k \in \{1,10\}$ .}
\end{figure*}

Subsequent change to the Hill coefficient with $n\geq 1$ shows the dynamic behaviour as in \fig{10}, where the control is also able to stabilise the system of oscillators simulating collaborative binding. A mixture of Hill coefficient values will result in other stable and unstable patterns emerging from the dynamics, depending on the ability of the control functions to remain the system in a periodic cycle or steady state.

The array of ARCC controlled BZ models shows that allostery can enhance the dynamic responses of the oscillators and return an unstable perturbed system to a stable state, or period. The ARCC method allows the direct relation between allostery and control of the underlying dynamic system to be studied and simulated dynamically.

\section{Discussion}
The necessity to replace the essential Michaelis-Menten (MM) model with a different expression, may not initially be very obvious. It should be recognised that the constraints imposed by the MM assumption is desirable only when there is no other method available to stabilise and map the rate response function. These constraints reduce the rate equations to a linearised approximation to ensure stability and accuracy in zero flux conditions alone \cite{Schuster.1999}.  It has become therefore essential to replace the model with a different approach that is not similarly constrained \cite{Savageau.1988}. Additionally, there is no real means of control within the standard MM approach, which leads to a systemic lack of functionality in the resulting model network. This then requires additional approaches to explain and analyse dynamic behaviour of metabolic systems, where the approximation is unable to model the biological complexities. 

The choice of the Rate Control of Chaos as a suitable expression to encapsulate kinetic reaction rates, is derived from its inherent assumption to limit the exponential rate of growth in chaotic systems. This derivation was originally inspired by a reaction rate equation but was, until now, not formally applied to this specific problem \cite{Scheper.2017}. As shown in \fig{1}, the RCC works to stabilise the nonlinear dynamics, and converges to a stable state readily. The generalisation of the RCC into ARCC via equations \eqref{eq:grcc1} and \eqref{eq:grcc2} showed that the method can be used consistently, but only without the MM approximation. The replication of the original results by Nielsen at al., shows how the ARCC method provides similar results as those produced experimentally, even without explicit tuning of the parameters to match the results.

The very large amount of work that has been produced to express countless experimental results into representative rate equations is directly applicable to the ARCC method. The Allosteric Rate Control of Chaos method enriches the ability to express reaction rates and allows a direct link between enzymatic control and substrate concentration to be established. The initial proportional approximation of the rate function is valid and contained within the ARCC control function. The resulting dynamic behaviour is in a controlled stable state or periodic orbit, and robust against perturbations and parameter changes. The dynamic response due to a change in a bifurcation parameter \fig{3} causes the control to adjust until a new stable state can be found. If the method was applied to wider metabolic models, this would explain how biological system maintain multiple stable states when perturbed and how external input does not cause the system to destabilise. The system aims to maintain stability of the local dynamics that allows the output (in this case NADH), to remain at a physiological level (\fig{6}).

The addition of the Arrhenius relation is, of course, of limited interest for a functioning physiological system, but demonstrates that the control can be used as a systemic adjustment to the entire reaction, changing the oscillation shape and frequency similar to temperature dependent systems. It should also be noted that body temperature is a widely used mechanism within biology to affect immune response and physiological developments. Additionally, the relation shows how the control itself may be adjusted by external parameters to achieve a specific response or desired throughput of the reaction cascade.

The question of allostery in relation to the MM condition has been confounding the elucidation of the relation of protein binding and reaction rate kinetics  to some extent \cite{Gunasekaran.2004,Nussinov.2015,Hofmeyr.1997ke}. Even though the relation seems explicit, the models based on Hill equations \cite{Hill.1977mw} and Michaelis-Menten are not readily reconcilable. The ARCC inclusion of both these two aspects as a means of allosteric control appears to resolve this conundrum, by creating a mechanism that allows control on the basis of substrate concentration, allostery and control adaptation by external mechanisms to be combined. Efforts that explore different values for the Hill coefficient, have shown that this may well be a variable under the influence of regulatory and external control mechanisms \cite{Ravera.2015}. The Belousov-Zhabotinsky models with ARCC control shows that this could well be a suitable approach to regulate the emerging dynamic behaviour of networks of reactions, to achieve a desired or required level of activity or optimise for a specific product or consumable.

The  generalised model of the ARCC as described by \eqref{eq:grcc1} and \eqref{eq:grcc2} can be extended in several different directions. The essential property of control that the method allows is maintained in the relations expressed in these equations but the ratio and shape of the underlying proportional representation may be better represented by more complex expressions. For example, the $\xi$ scalar used for the control function in \eqref{eq:grcc2} can be replaced by an entire expression that contains the biophysical dynamic behaviour of subunit binding, representing the state of the enzyme, and also allows compartmental and population dynamics of both reactants and enzymes to be expressed in more detailed manner without losing the essential control of the reaction. Changes in the bias term, the Hill coefficients, and the combination of reaction kinetics in the fractional expression are also strong candidates for control mechanisms on the basis of the biophysical properties of the enzyme and regulatory control.

The BZ array simulations also demonstrate the property of the ARCC method to stabilise and control reaction cascade dynamics. This will allow more insight into the issues surrounding local dynamic behaviour of pools of biochemical reactants disrupt and enhance other processes in the immediate neighbourhood. It provides a robust mechanism that ensures stable dynamic response to continual external perturbation and competition for molecular binding. Large scale simulation of interacting reaction dynamics will therefore become feasible to study the physiological behaviour in more detail and allow potential disorder and treatment approaches to be identified and addressed.

\section{Conclusion}
The Rate Control of Chaos method is capable of controlling nonlinear and chaotic dynamic systems. The generalised Allosteric Rate Control of Chaos method can be applied to control biochemical reaction rate kinetics, reproducing the essential properties of the Monod-Wyman-Changeux kinetic models that are based on Michaelis-Menten. ARCC is also robust when externally perturbed and can maintain a stable state across a wide range of model parameter values. The cardinal property of ARCC as a control method ensures that the system under control does not readily destabilise and provides a means for biochemical control mechanisms to exert direct control on the reactions. The effect of allostery on the ARCC system shows that both competitive and collaborative binding can affect the dynamic stability and response of the reaction cascade even under varying conditions. The ARCC model allows an entirely novel approach to understanding biochemical kinetics, creating direct relations with the mechanistic properties of enzyme reaction complex, and allows in depth study of the biophysical complex with the resulting kinetic dynamics.\\

\noindent The author wishes to declare no competing interests, and that the work was performed without funding.
\pagebreak


\end{document}